\documentclass{emulateapj}

\begin{document}

\shorttitle{Model Mapping of PDR NGC 2023-South}
\shortauthors{Sheffer et al.}

\title{PDR Model Mapping of Physical Conditions via \textit{Spitzer}-IRS
Spectroscopy of H$_2$: Theoretical Success toward NGC 2023-South}

\author{
Y. Sheffer\altaffilmark{1},
M. G. Wolfire\altaffilmark{1},
D. J. Hollenbach\altaffilmark{2},
M. J. Kaufman\altaffilmark{3},
and M. Cordier\altaffilmark{3}
}

\altaffiltext{1}{Department of Astronomy, University of Maryland, College Park, MD 20742, USA; ysheffer@astro.umd.edu}
\altaffiltext{2}{SETI Institute, 189 Bernardo Ave, Mountain View, CA 94043, USA}
\altaffiltext{3}{Department of Physics and Astronomy, San Jose State University, One Washington Square, San Jose, CA 95192, USA}

\begin{abstract}

We use the IRS on \textit{Spitzer} to observe the southern part of the
reflection nebula NGC 2023, including the Southern Ridge, which is a
photodissociation region \textit{par excellence} excited by HD 37903.
Five pure-rotational H$_2$ emission lines are detected
and mapped over and around the Southern Ridge in order to compare with
predicted level column densities from theoretical PDR models.
We find very good agreement between PDR model predictions and emission line
intensities and ratios measured with \textit{Spitzer}, leading us to conclude
that grain photoelectric heating sufficiently warms the gas to produce the
observed ${\rm H_2}$ line emission via collisional excitation.
On the Southern Ridge, we infer a hydrogen nucleus density
$n_{\rm H} \approx 2 \times 10^{5}$ cm$^{-3}$ and radiation field strength
$\chi \approx 10^{4}$ relative to the local Galactic interstellar
radiation field.
This high value for $\chi$ independently predicts a distance toward HD 37903
of 300 pc, and is consistent with the most recent \textit{HIPPARCOS} results.
Over the map we find that both $n_{\rm H}$ and $\chi$ vary by a factor of
$\sim$3.
Such 2-D variations provide clues about the underlying 3-D structure
of the Southern Ridge field, which appears to be the tip of a molecular cloud. 
We also map variations in excitation temperature and the ortho-to-para ratio,
the latter attaining values of $\sim$1.5---2.0 on the Southern Ridge, and find
that PDR modeling can readily reproduce observed ortho-to-para ratios that
are $<$3 for rotational excitation dominated by collisional processes.
Last, the stars Sellgren C and G are discovered to be resolved on archival
\textit{HST} images into two point sources each, with separations
of $\lesssim$0$\farcs$5.

\end{abstract}


\section{Introduction}

Photodissociation regions (PDRs) are regions in interstellar clouds in which
far-ultraviolet ($6 < h\nu < 13.6$  eV; FUV) radiation plays a significant
role in the heating and/or chemistry \citep{Tielens1985}.
For example, PDRs are found in reflection nebulae and molecular cloud surfaces,
where the radiation from nearby OB stars illuminates the clouds.
The incident starlight is absorbed by dust and polycyclic 
aromatic hydrocarbons (PAHs) and is mostly used to excite the PAHs and 
heat the grains.
However, a fraction ($\sim$0.1---1\%) of the absorbed FUV starlight 
may be converted into energetic photoelectrons, which are ejected from PAHs 
and grains, and subsequently heat the gas.
The strong FUV radiation acts as a beacon to illuminate the cloud structure,
and to photodissociate, ionize, and excite gas phase chemical species,
which otherwise would not be seen in emission.
Thus, PDRs emit
strong far-infrared continuum emission from grains, as well as infrared, 
submillimeter, and millimeter-wave line emission arising from the warm gas. 
The FUV radiation can affect the chemistry in molecular clouds to a depth of  
$A_V \sim 5$ by maintaining the oxygen that is not tied up in CO
in atomic form \citep{Tielens1985,Hollenbach2009}.
This depth is comparable to the mean column density in
giant molecular clouds \citep{Solomon1987,Heyer2009}.
The same PDR physics that is at work at the surfaces 
of molecular clouds also acts in the diffuse interstellar medium 
\citep{Wolfire2003} and, therefore, much of the ISM is found in PDRs. 

In general, the theoretical models \citep[e.g.,][]{Sternberg1995,
Wolfire2003,Kaufman2006,LePetit2006}
do a good job in predicting the atomic fine-structure
line intensities and line ratios in PDRs; however, several observations
of line emission from H$_2$ using \textit{ISO} \citep{Timmermann1996,
Fuente1999,Draine2000} and \textit{Spitzer} \citep{Goldsmith2010}
seem to indicate temperatures higher than those predicted by grain
photoelectric heating alone. 
In addition, observations and modeling of high-$J$ lines of CO
\citep{Jaffe1990,Steiman1997} would also
indicate that models underestimate the gas temperature ($T_{\rm gas}$).
\cite{Habart2011} found order-of-magnitude discrepancies
between their PDR model results and \textit{Spitzer} H$_2$ data
for rotational levels $J \ge 3$ toward mainly low-excitation PDRs.
\cite{Dedes2010} were able to fit observations of high-$J$ CO lines by using
spherical PDR models for an ensemble of clumps distributed in size and mass.
\cite{Weingartner1999} have suggested that radiation forces on
grains increase the dust/gas ratio in PDRs thus leading to enhanced
heating rates.
Dissipation of turbulence \citep{Falgarone2005} and shocks \citep{Habart2011}
might also be an important source of heating in
low-FUV field environments such as the Taurus molecular cloud
\citep{Goldsmith2010} or in the diffuse ISM.

Molecular hydrogen, H$_2$, is a sensitive probe of PDRs in our Galaxy
\citep{Allers2005} as well as in highly redshifted galaxies \citep{Sheffer2009}.
The pure-rotational ($v$ = 0) transitions of H$_2$ in the mid-IR are
readily observed by space instruments, such as the IRS on \textit{Spitzer}
\citep{Houck2004}. 
The high spatial resolution ($<$10$\arcsec$) of our \textit{Spitzer}
observations more clearly isolate the emission structures and different
physical regions compared to the previous \textit{ISO} observations. 
In general, transitions between low-$J$ levels of the $v$ = 0 state probe
$T_{\rm gas}$ owing to their low critical densities, while
the higher-excitation energy levels are pumped by FUV radiation and probe the
radiation field strength and gas density \citep{Sternberg1989,Burton1990}.

This project has two main goals: first, to use H$_2$ observations
from \textit{Spitzer}-IRS of the reflection nebula \object{NGC 2023} to 
derive average gas physical conditions in the warm molecular regions and,
second, to assess the reliability of PDR models to model the gas emission. 
In $\S$ 2 we track the changing values for the estimated
distance of NGC 2023, present
recent imaging results of the nebula, and describe our spectral
mapping of the Southern Ridge (a.k.a. filament or bar, hereafter SR)
area in NGC 2023 based on \textit{Spitzer} data collected by the IRS.
The intensities of pure-rotational emission lines from H$_2$
are measured and converted into level column densities.
These observables are compared in $\S$ 3 with model predictions
based on updated \cite{Kaufman2006} models to show that a high level of
agreement exists between theory and observation.
Model parametrization allows us to map the 2-D distribution of total gas
density and FUV strength, and to glimpse the 3-D structure
of the nebula based on model fits.
Section 3 also presents maps of the ortho- to para-H$_2$ (o- to p-H$_2$)
ratio (OPR) and of excitation temperatures ($T_{\rm ex}$) for H$_2$, and
concludes with a discussion of detected
fine-structure line emission from \ion{Si}{2} and \ion{Fe}{2}.
Section 4 is dedicated to the role of photoelectric 
heating compared to other processes.
Finally, a concluding section provides a textual closure for the paper
by emphasizing our main results.

\section{The Target and the Data}

\subsection{What is the Distance to NGC 2023?}

As a reflection nebula surrounding the hot B1.5~V star \object{HD 37903}, the
issue of the distance of NGC 2023 is intimately tied to the distance of its
central star, as well as being an essential ingredient in estimating the
physical separation between HD 37903 and the SR.
Most of the studies dealing with NGC 2023 over the last 30 years
have employed a consistent range of values of 450---500 pc for the distance
toward this object, as well as toward the entire nebular complex in Orion.
However, assuming the depth is roughly the same as the projected extent,
the angular size of the constellation
of Orion predicts a depth to distance ratio of $\sim$35\%, amounting to a
range of $\sim$140 pc for a central value of 400 pc.
Indeed, \cite{Anthony-Twarog1982} derived a statistical distance toward
Orion B stars of 380$^{+100}_{-80}$ pc.
This shorter distance scale was employed by the \cite{Wyrowski1997} and
\cite{Martini1999} studies of NGC 2023, but it was quickly abandoned in favor
of the longer one based on the \textit{HIPPARCOS} parallax of 
2.12 $\pm$ 1.23 mas \citep{Perryman1997},
placing HD 37903 at a rather imprecise distance of 470$^{+650}_{-170}$ pc.

However, very accurate parallax measurements with very long baseline
interferometry show the distance to the Orion Nebula to be 400 pc, with
uncertainty of 2---6\% \citep{Sandstrom2007,Menten2007}.
Additional support for a shorter scale is provided by \cite{Caballero2008},
who derived a distance of 334$^{+25}_{-22}$ (or, less likely, 385 $\pm$ 15) pc
toward $\sigma$ Ori, a star in the angular vicinity of HD 37903.
Moreover, the cloud complex that includes NGC 2023 and the
Horsehead Nebula appears to be in front of $\sigma$ Ori
\citep{Mookerjea2009}.
We shall adopt 350 $\pm$ 50 pc as the probable distance toward HD 37903
as suggested by \cite{Mookerjea2009}.
Clearly, HD 37903 is located at the very near side of the Orion nebular
complex\footnotemark[4], implying that the SR is separated from the star by
a projected distance of (4.0 $\pm$ 0.6) $\times 10^{17}$ cm, or
0.13 $\pm$ 0.02 pc.

\footnotetext[4]{The short distance scale has now been confirmed by
\textit{HIPPARCOS}, see end of $\S$ 5 for details.}

\subsection{Non-\textit{Spitzer} Images}

\begin{figure}
\epsscale{0.9}
\plottwo{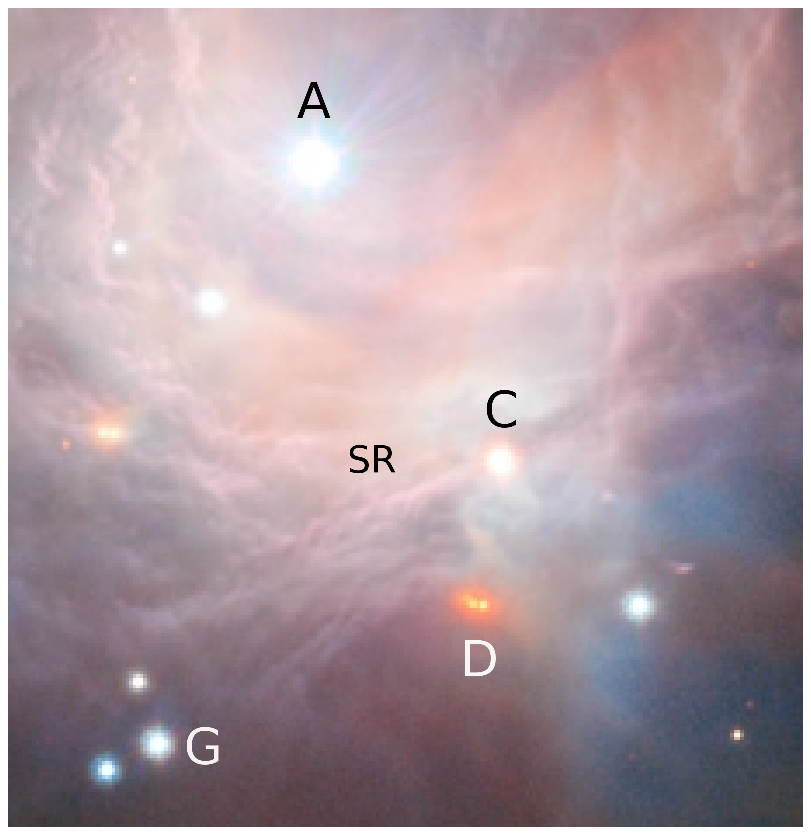}{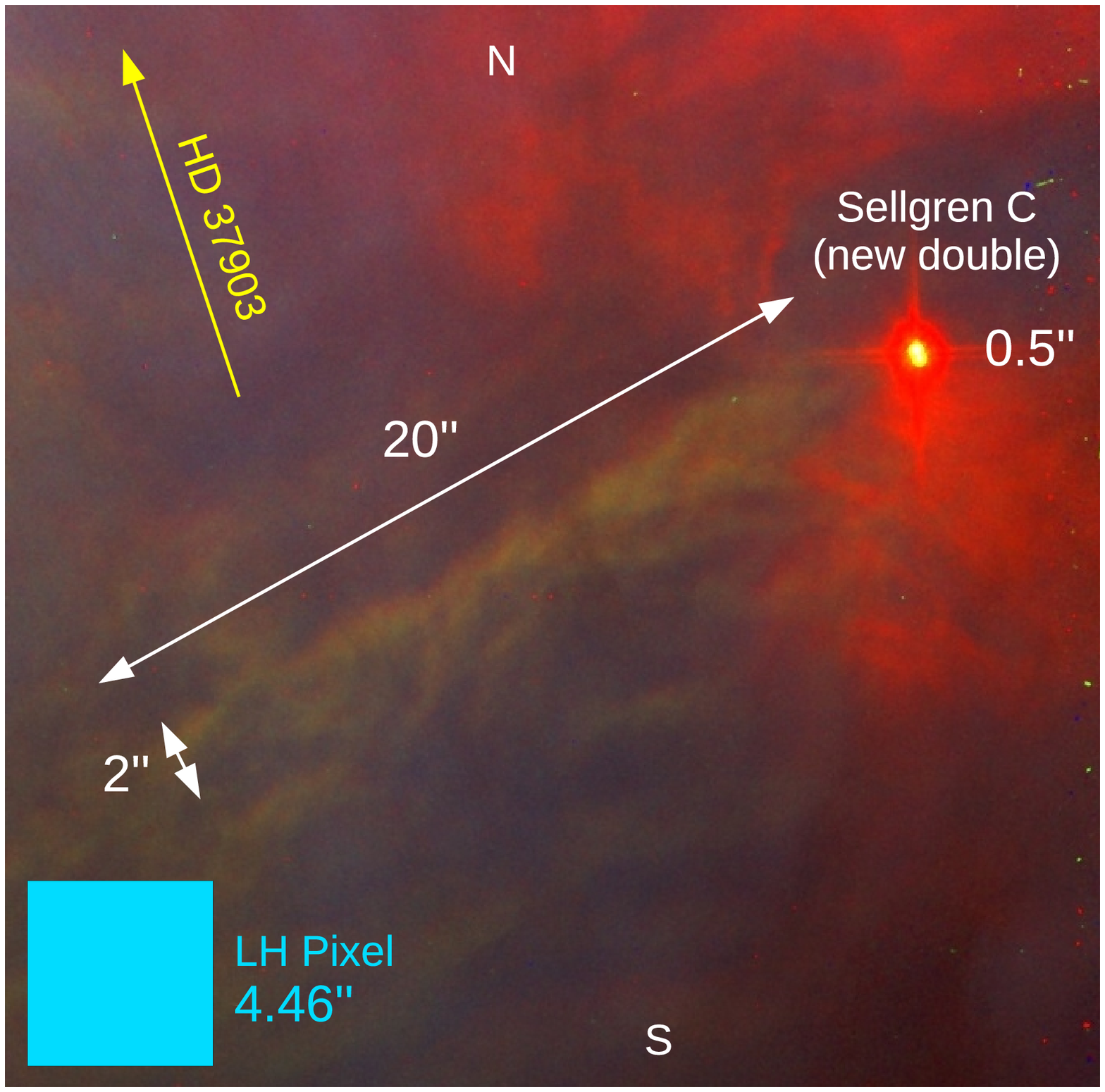}
\caption{Left panel shows a near-infrared view of the southern half of NGC 2023
as combined from J (``blue''), H (``green''), and K (``red'') exposures.
This highly magnified region of size $\approx 3\arcmin \times 3\arcmin$ shows
HD 37903 (Sellgren A) as the brightest star toward the top, and
a pink-colored Southern Ridge (SR) just below the center of the image.
Additional stars are identified by their Sellgren letters.
Credit: ESO/J. Emerson/VISTA/Cambridge Astronomical Survey Unit
(ESO release 0949).
Right panel shows a magnified region of $\approx 30\arcsec \times 30\arcsec$
from an \textit{HST}-ACS image, providing
the highest angular resolution view of the SR to date.
A blue square shows the size of an LH pixel from \textit{Spitzer}-IRS.
All figure labels were inserted manually and should not be presumed to have
a level of precision better than 10\%.
Credit: http://en.wikipedia.org/wiki/NGC\_2023, based on ACS data set j8mw01.}
\label{fig:show_2023}
\end{figure}

In order to become familiar with the appearance and structure of NGC 2023,
we include here two high-quality public data products
that have been obtained in recent years.
The first is a near-IR image obtained at ESO (left panel of
Figure~\ref{fig:show_2023}),
which clearly shows the structure of the nebula that heretofore
had been hidden from visual view.
The star HD 37903 has carved a quasi-spherical cavity into the dense molecular
material out of which it formed.
Consequently, its energetic FUV radiation illuminates the ridges of
high-density gas, producing a reflection nebula and emission ridges that
are detected in H$_2$ line and dust (as well as PAH) continuum emission.
Sellgren D, the eye-catching orange object below the pink SR,
is resolved into at least three sources, whereas it was
seen as an elongated object in previous renditions.

A portion of an archival image from the \textit{Hubble Space Telescope} is
also shown in Figure~\ref{fig:show_2023}.
Taken in red visible light with the ACS, it is the highest-resolution image
ever taken of NGC 2023.
The SR is resolved into an ensemble of clumps, all part of massive ridges
of cloud tops that are basking in the FUV starshine of HD 37903.
The SR was already known from ground-based work to be $\sim$2$\arcsec$ wide
\citep{Field1994,Field1998},
but the newly resolved clumps are smaller than that by a factor of a few.
They are, on the other hand, much larger than the 0$\farcs$05 angular size
of each pixel in the ACS field of view.

As a bonus, the (full) \textit{HST} image also resolves two stars into their
constituent components for the first time.
These are Sellgren C near the western end of the SR, and Sellgren G, which is
labeled on the ESO image.
However, we cannot be certain that two actual binaries are involved, since
young stellar objects, such as these PMS stars \citep{Sellgren1983}, may be
surrounded by a residual
edge-on debris disks from the formation process, which might mimic a binary
source.
According to \cite{DePoy1990}, their star 9 (Sellgren G)
is consistent at the 1-$\sigma$ level with standard reddening vectors, i.e.,
it is not a convincing case of special extinction by a PMS ``cocoon''.
In both cases the components are 0$\farcs$4 $\pm$ 0$\farcs$1 apart,
or projected separations of $\sim$150 AU.

\subsection{\textit{Spitzer} Spectroscopy}

We acquired \textit{Spitzer}-IRS spectra of H$_2$ emission lines toward
NGC 2023 by employing the three modules SL, SH, and LH, as listed in Table 1,
where the second letter `L' or `H' stands for low- or high-spectral resolution,
respectively.
The lower-resolution SL module is split into orders SL1 and SL2 and provides
potential coverage of H$_2$ emission lines from $J$ = 4---9 levels, i.e., of
transitions S(2) through S(7).
However, owing to substantial blending with strong PAH features, we could
detect only S(5) at 6.91 $\micron$ in SL2 data, and S(3) and S(2) at 9.66
and 12.28 $\micron$, respectively, in SL1 spectra.
SH also covers S(2), but at resolution $\approx$10 times higher than that of SL,
and has exclusive coverage of S(1) from $J$ = 3 at 17.03 $\micron$,
whereas LH provides the only coverage of S(0) from $J$ = 2 at
28.22 $\micron$.

\begin{figure}
\epsscale{0.9}
\plotone{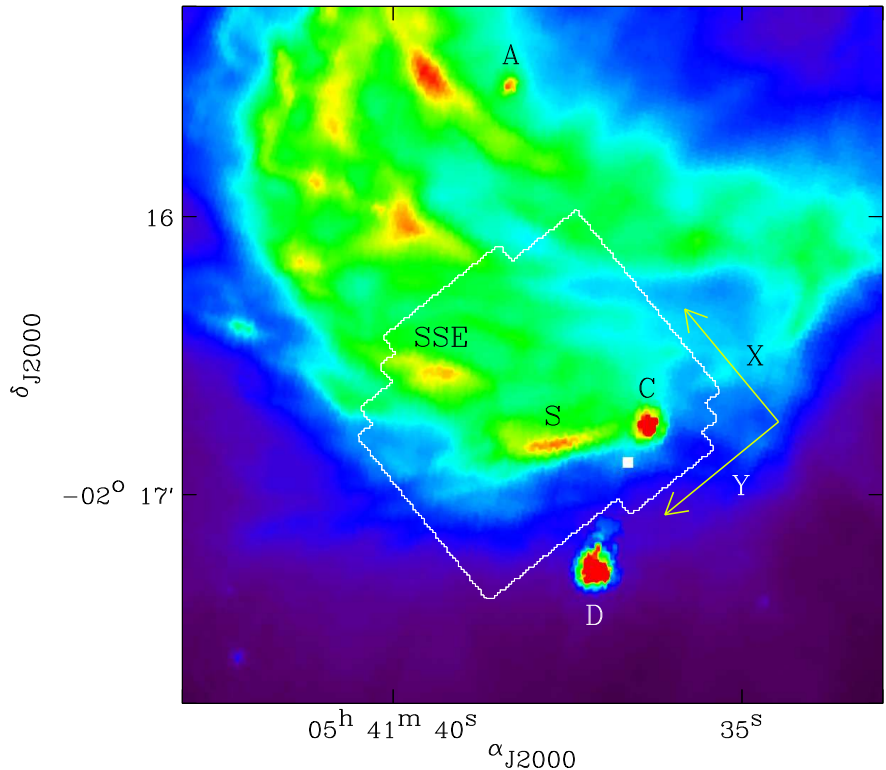}
\caption{IRS all-module common field (irregular outline) overlaying the IRAC
channel 4 image of NGC 2023, which is dominated by PAH emission at 8 $\micron$.
All maps to follow shall employ the instrumental orientation (vectors X, Y).
As in left panel of Figure~\ref{fig:show_2023}, Sellgren A is the IR-faint
HD 37903, whereas Sellgren C and D are two IR-bright young stellar objects.
S is the SR, a narrow H$_2$ emission filament, whereas SSE is
the South-Southeastern Ridge, a wider emission clump.}
\label{fig:show_irac}
\end{figure}

The data were processed into cubes using the matching 
\textit{Spitzer} Science Center pipeline version (S18.7.0) and
CUBISM \citep{Smith2007} version 1.7.
All non-LH exposures were spatially degraded by re-gridding onto the (larger)
pixels of the LH field of view prior to further analysis.
Figure~\ref{fig:show_irac} shows the proper celestial location and orientation
of the LH field over an 8 $\micron$ image from \textit{Spitzer}-IRAC.
With 15 $\times$ 14 pixels in the LH field, each 4$\farcs$46 across, the
irregular border of the combined intersection area of all modules measures
$\approx 1\arcmin \times 1\arcmin$
along the instrumental X and Y directions.

\begin{figure}
\epsscale{0.6}
\plotone{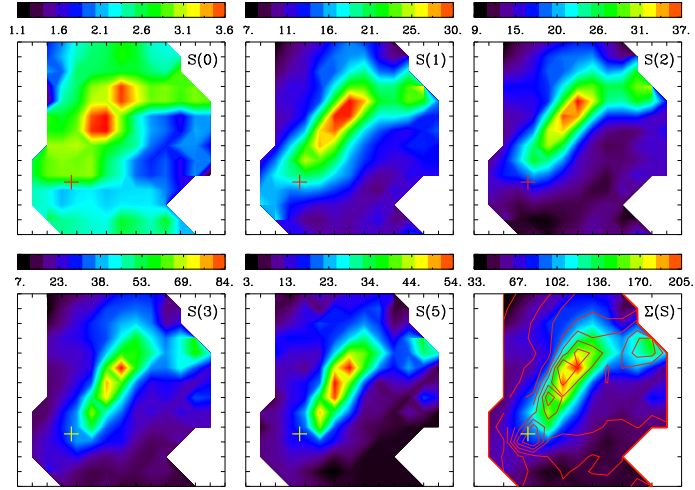}
\caption{Intensity maps of H$_2$ line emission (uncorrected for extinction)
toward the SR of NGC 2023.
The color bars show the intensity scale in units of
10$^{-5}$ erg s$^{-1}$ cm$^{-2}$ sr$^{-1}$.
Each box spans 14 $\times$ 13 pixels, or $\approx 1\arcmin \times 1\arcmin$.
First five panels from top left show detected transitions from $J$ = 2, 3, 4, 5
and 7 in the observed LH frame.
Last panel employs the total intensity of all five emission lines as a
background for contours of continuum intensity from the 8 $\micron$ IRAC image
shown in Figure~\ref{fig:show_irac}.
The `+' indicates the position of Sellgren C.}
\label{fig:show_5xem}
\end{figure}

Emission line maps were constructed by using IDL/GAUSSFIT to fit line
profiles and to derive integrated line intensities that included continuum
fitting and removal.
Table 2 lists derived GAUSSFIT parameters and their uncertainties for the
case of map medians (dominated by off-SR pixels) and for the
case of the single on-SR pixel LH[7:8], which coincides with the location of
peak H$_2$ emission.
Figure~\ref{fig:show_5xem} shows the spatial distribution
of H$_2$ emission for the five detected transitions S(0), S(1), S(2), S(3),
and S(5).
In these maps, the SR is a very prominent source of H$_2$
line emission, with a partial coverage of additional emission from the
South-Southeastern Ridge toward the instrumental top right corner.
(Note that here, and in following Figures, the instrumental orientation of the
LH map is to be employed in order to avoid both additional interpolation of the
data and the wasteful white margins inherent in celestial orientation.)
The H$_2$ shows good agreement in position and orientation with the SR
seen in the IRAC 8$\micron$ image, which traces mainly PAH emission
(Figure~\ref{fig:show_irac}).

\begin{figure}
\epsscale{0.5}
\plotone{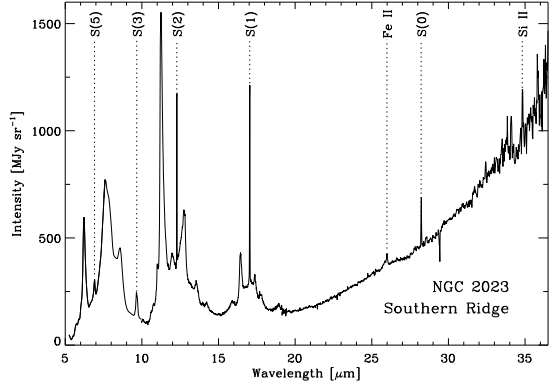}
\caption{Full spectral coverage from the four IRS modules SL2, SL1, SH, and LH
toward NGC 2023, as obtained by averaging 15 pixels that sample SR emission.
H$_2$ and atomic emission lines are identified, and their rest wavelengths
are indicated by dotted lines.
Strong PAH features and dust continuum are evident.}
\label{fig:full_spec}
\end{figure}

The target area observed by \textit{Spitzer} toward NGC 2023 shows prominent
and broad emission features from PAH molecules together with strong and narrow
(unresolved) emission lines from H$_2$, see Figure~\ref{fig:full_spec}.
This spectrum was obtained by averaging 15 LH pixels that sample the emission
from the SR in all four IRS modules.
Two minor contributions from atomic species on the SR belong to
[\ion{Fe}{2}] and [\ion{Si}{2}], at 25.99 and 34.82 $\micron$, respectively,
which are expected to be PDR observables \citep{Tielens1985,Kaufman2006}
and are presented in $\S$ 3.4.
Two other fine-structure transitions commonly found in H II spectra,
[\ion{Ne}{2}] at 12.81 $\micron$ and [\ion{S}{3}] at 18.71 $\micron$,
are detected at extremely weak levels of emission all over the map.
We shall concentrate in this analysis on the H$_2$ lines, comparing them
with model predictions.
The analysis of the PAH features shall be presented in a later paper 
(Peeters, et al.\ 2011, in preparation).

\subsection{IRS Calibration Issues}

\subsubsection{Differential Extinction}

We applied extinction corrections to our data consistent with previous
studies of H$_2$ line emission from NGC 2023.
\cite{Burton1993} found that $A_K$ = 0.3 mag from their vibrationally excited
emission lines of H$_2$, yielding $A_V$ = 2.8---2.3 mag for the range
$R_V$ = 3.1---5.5 \citep{Mathis1990}, whereas \cite{Burton1998}
mentioned that $A_V$ is likely to be $\sim$3---5 mag, or
$A_K$ $\sim$ 0.3---0.65 mag.
\cite{Draine1996} and \cite{Draine2000} adopted
$A_K$ of 0.2 and 0.5 mag, respectively.
For our corrections we adopted $A_K$ = 0.5 mag, which corresponds to
$A_V$ = 3.8---4.7 mag.

Using the extinction curve of \cite{Mathis1990}, the extinction correction
varies between 6---29\%, with the largest correction applying to S(3).
The latter value is similar in magnitude to systematic effects
in the absolute calibration of \textit{Spitzer}-IRS fluxes, estimated to be
$\sim$20---25\% \citep{Galliano2008,Dale2009}.
For the rotational levels $J$ = 2 and 5 with the lowest and highest extinction
corrections, respectively, we determined that varying both $N_J$ values by
$\pm$20\% shifts their derived $T_{52}$ value (i.e., $T_{\rm ex}$ of level
$J$ = 5 relative to level $J$ = 2) by +3 and $-$8\%.
This smaller change, owing to a weak dependence on the ratio of column
densities (since $T_{52} \propto {\rm ln}[N_2/N_5]$),
is consistent with \cite{Dale2009}, who reported $\pm$10\%
for derived line ratios extracted from \textit{Spitzer}-IRS data.

\subsubsection{Background Subtraction}

Background contribution from zodiacal emission is always present toward
celestial targets, as a function of space (direction) and time (date).
Toward NGC 2023, dedicated background exposures were taken only for the SL
observations, revealing a wavelength-averaged ratio of background to data
of 10\%.
We employed the zodiacal emission calculator from
SPOT, for each date and pointing center of each observation
in order to compute the expected level of background emission for our data,
and found good agreement between observed and calculated zodiacal emission
for the SL data.
Thus we are assured that subtraction of calculated zodiacal emission from data
that lack background exposures does not lead to any significant errors.
Furthermore, since our analysis concerns emission intensity following continuum
subtraction, such intensities are insensitive to the
presence of continuum-like background levels.

\subsubsection{Intensity Inter-Calibration}

There is a significant wavelength overlap at 9.97---14.74 $\micron$
between the SH and SL1 modules.
Previous works have shown a continuum mismatch between modules, with different
modules requiring different scale factors to bring the continuum into agreement.
For example, \cite{Brandl2004}, had to shift their SH data by +36\%
and their SL1 data by +17\% relative to LL data, showing that in their case,
the SL1 scale was higher by +16\% than the scale of the SH module.
\cite{Quanz2007}, on the other hand, found higher readings from SH
data relative to SL1 data, and attributed these 8---25\% scale shifts
to different slit orientations relative to source emission.
\cite{Beirao2008} found up to 50\% differences in fits of PAH features
based on SH and SL1 spectra.

We performed a pixel-wise $\chi^2$ analysis of SH data versus SL1 data,
which included background subtraction as well as SH resolution degradation
to the lower spectral resolution of SL1 data.
The absolute intensity scale of the SH data was found to be higher by 19\%
than that of SL1 data.
Since we consider the higher signal-to-noise SL1 data to be more reliable
spectrophotometrically, the SH (and LH) data were re-scaled by 0.84 prior to
measuring integrated intensity values for the S(0), S(1), and S(2) lines.

The combined uncertainty in our measurements owing to the dominant calibration
issues of absolute flux uncertainty and inter-module uncertainty is
$\lesssim$30\%.

\subsection{Comparison of \textit{Spitzer} IRS with \textit{ISO} SWS}

The extinction-corrected H$_2$ line intensities were converted into level
column densities using $N_J = 4 \pi I_J / A_J \Delta E_J$ cm$^{-2}$, where
$N_J, I_J, A_J$, and $\Delta E_J$
stand for the column density, emission intensity, Einstein A-coefficient, and
transition energy for each upper level $J$.
The H$_2$ rotational emission lines are quadrupole transitions and thus an
optically thin conversion is appropriate for all environments in the ISM.
For mapping purposes, this procedure was followed for each pixel over
the field of view.
To compare with previous ISO observations, we employ the mean on-SR 
H$_2$ column density based on $I_J$ values from 15 pixels that sample SR
emission.
These pixels were also employed in the extraction of the average on-SR spectrum
shown in Figure~\ref{fig:full_spec}.
The SR-averaged log $N_J$ values for $J$ = 2, 3, 4, 5, and 7 are 20.29, 19.77,
18.95, 18.60, and 17.48 cm$^{-2}$, respectively.

\textit{ISO}-SWS observations of pure rotational transitions toward the SR
of NGC 2023 were shown in Figure 5 of \cite{Draine2000}, from which
we extracted $N_J$ values for the five
emission lines detected in our \textit{Spitzer} data.
These values were then rescaled by 1/1.8 in order
to remove the arbitrary beam filling factor employed by \cite{Draine2000},
resulting in log $N_J$ values of 20.44, 19.84, 18.90, 18.56, and 17.24
cm$^{-2}$.
Consequently we find \textit{ISO} to \textit{Spitzer} column density ratios of
1.4, 1.2, 0.9, 0.9, and 0.6, averaging 1.0 $\pm$ 0.3, which has the
uncertainty expected for two data sets with $\sim$20\% uncertainty each.
Still, it is fascinating that the ratio appears to have a monotonic
decline with increasing $J$, possibly related to differences in beam sizes
employed for differing wavelength regimes on both spacecraft.

\cite{Habart2004} quote the (extinction-uncorrected) \textit{ISO} observable
$I = 1.65 \times 10^{-5}$ erg s$^{-1}$ cm$^{-2}$ sr$^{-1}$
for the S(3) line, taken with a
19$\arcsec$ beam centered on a point 60$\arcsec$ due south of HD 37903.
This position lies north of the SR and prevents us from directly comparing it
with SR intensity values, however it still lies within our mapped region.
We simulated this beam on our IRS map and found that the
\textit{Spitzer} H$_2$ emission is 42\% higher than
the value given by \cite{Habart2004},
hence a column density ratio of \textit{ISO}/\textit{Spitzer} = 0.7,
which is consistent with the ratios given above.
The agreement between IRS and SWS results is quite good considering the
uncertainty in the \textit{ISO} beam
filling factors and extended source calibrations for both \textit{ISO} and
\textit{Spitzer}.

\section{Modeling \textit{Spitzer} Data with PDR Models}

\subsection{Model Parameters}

In order to test the results of PDR models, we compare $N_J$(H$_2$) values
derived from our observed emission lines toward NGC 2023 with values predicted
by our PDR model \citep{Kaufman1999,Kaufman2006}.
This model includes the calculation of H$_2$ processes from the 
\cite{LePetit2006} code as discussed in \cite{Kaufman2006}. 
These include radiative excitation and dissociation, dissociation heating, 
collisional excitation and deexcitation, radiative cooling, and heating by
deexcitation of excited levels.
The o-H$_2$ to p-H$_2$ conversion on grains is included in the
\cite{LePetit2006} code as described in \cite{LeBourlot2000}.
We also include a factor of 2 enhanced H$_2$ formation rate as suggested by
\cite{Habart2004} and discussed in \cite{Kaufman2006}.
The OH and CO chemistry has been updated as in \cite{Wolfire2010} and we 
test models for normally-incident photons.
The model output consists of the H$_2$ column densities in each $J$
level integrated along the normal to the PDR surface, as well as the normally
emitted H$_2$ line intensities.
The ro-vibrational quadrupole transitions are all optically thin.
We will vary two main model parameters to obtain a best fit to the observations
while holding all others constant.
These are the hydrogen nucleus density, $n_{\rm H}$, and the incident radiation
field, $\chi$, in units of the \cite{Draine1978} interstellar
radiation field.
We use the notation that $\chi$ is the ratio of FUV field incident on the
surface of the PDR divided by the free-space field in the local interstellar
radiation field.
Thus $\chi = F_{FUV}/4 \pi I_D$, where $F_{FUV}$ is the incident FUV flux and
$I_D = 2.2 \times 10^{-4}$ erg s$^{-1}$ cm$^{-2}$ sr$^{-1}$ is the Draine
intensity for the local ISM.

In order to establish initial parameter values for $n_{\rm H}$ and $\chi$,
as well limit their variation during modeling to values that are consistent
with known ranges of other observables, we need independent methods for
estimating $n_{\rm H}$ and $\chi$.
\cite{Wyrowski2000} applied PDR model results to their observed angular
separations between emission from H$_2$ and C91$\alpha$ toward NGC 2023.
They found a range of $n_{\rm H}$ = 0.6---1.4 $\times 10^5$ cm$^{-3}$,
which agrees with
previous findings that this is a relatively high-density PDR and provides
us with an initial density value of $n_{\rm H} \approx 10^5$ cm$^{-3}$.
The FUV field reaching the PDR gas from HD 37903 depends on the latter's
luminosity and the projected (lower limit) physical separation between the two,
which depends on the distance of NGC 2023 from Earth.
For a B1.5~V star of 12 $M_{\Sun}$ \citep{Conti2008} we find an FUV
luminosity of $1.13 \times 10^4 L_{\sun}$, based on \cite{Parravano2003}.
An angular separation of 78$\arcsec$ between the star and the SR results
in an FUV field strength of $\chi = 9.27 \times 10^8 / D^2$, where
D is the distance in pc to HD 37903.
Using the adopted value of $D$ = 350 $\pm$ 50 pc ($\S$ 2.1) predicts
a range of fairly high values for $\chi$ of $7.6^{+2.7}_{-1.8} \times 10^3$.
We shall explore a range of $\chi$ values corresponding to
$\sim$2-$\sigma$ variation around this central value.

In comparing normal model line intensities to observations there are several
factors to keep in mind, which can be divided into two classes.
One class includes factors that may increase the ratio of observed
intensity to model intensity.
The first factor is the possible presence of multiple PDRs along
the line of sight, $f_{\rm P}$.
A second factor is the inclination angle ($\theta$) of a single
PDR layer relative to the line of sight.
Limb brightening is expected to occur for any PDR inclination
$\theta > 0\degr$, causing the source line intensity
to appear brighter than a face-on ($\theta \equiv 0\degr$) model PDR by the
factor $f_\theta$ = 1/cos($\theta$).
Both factors can be lumped together as a single raising factor,
$f_+ = f_{\rm P} f_\theta \geq 1$.

The other class to consider includes factors that may decrease the ratio of
observed intensity to model intensity.
First is the fraction of the beam area that is filled by the source emission,
$f_{\rm B}$.
A second factor involves the incidence angle $\phi$ of FUV illumination,
which affects the penetration depth of the radiation field.
The intrinsic (deprojected) geometry of the exciting star (HD 37903) to the SR
should produce a range of values between strictly normal ($\phi \equiv 0\degr$)
and strictly parallel ($\phi \equiv 90\degr$) incidence,
depending on the details of the shapes of gas clumps on the SR.
For any value of $\phi > 0\degr$, the intensity of observed H$_2$ emission is
cut down by cos($\phi$) owing to the oblique path of the FUV photons through
the gas layer,
and thus we have for the lowering factor $f_- = f_{\rm B} f_\phi \leq 1$.

The effects of the raising and lowering factors obviously work in opposite
directions, such that the total effective ratio of data to model is
$f_{\rm eff}$ = $f_+ f_-$ = $f_{\rm P} f_\theta f_{\rm B} f_\phi$.
Effective ratios for map pixels will be derived as logarithmic differences
(or model "shifts") in $\S$ 3.2 via
matching of absolute model $N_J$ values between models and data.
The value of $f_{\rm eff}$ only tells us whether $f_+$ or 1/$f_-$ is the
dominant effect, but not the actual values of any individual factors involved.

\subsection{Absolute Column Densities as a Function of $J$}

\subsubsection{A Single Normal Model}

\begin{figure}
\epsscale{0.5}
\plotone{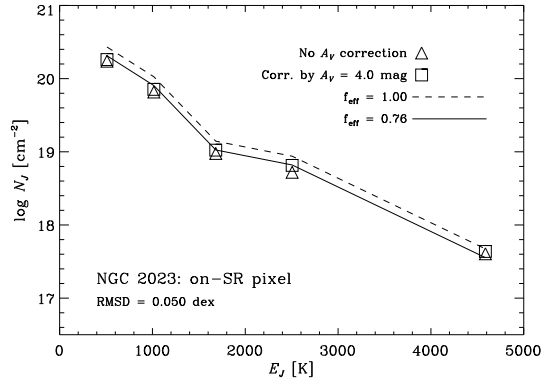}
\caption{Observed $N_J$ values (symbols) for the single on-SR
pixel LH[7:8] compared with model results (lines).
Boxes show $\Delta$$N_J$ = $\pm$20\% corrected for $A_V$ = 4.0 mag and
triangles show the data prior to extinction correction.
The dashed line shows the unshifted log $n_{\rm H}$ = 5.3 and
$\chi = 10^4$ model.
The solid line shows the same model shifted by $-$0.12 dex, or
$f_{\rm eff}$ = 0.76 (see text), following RMSD minimization.}
\label{fig:pix_vs_mod}
\end{figure}

The absolute column density values, $N_J$(H$_2$), constitute our primary means
of comparing PDR model results with the observations.
Figure~\ref{fig:pix_vs_mod} illustrates such a comparison for a single on-SR
pixel, where a run of observed $N_J$ values vs $E_J$, the level energy above
the ground state, is compared with output from the best matching model.
For illustration purposes, data are shown both before and after extinction
correction is applied.
The best match between models and data was determined from the smallest
root mean square deviation (RMSD) of the differences
in dex between modeled and observed absolute $N_J$ values.
In this way, both relative $N_J$ ratios between even- and odd-$J$ levels,
as well as global absolute $N_J$ values, were being fitted with the
best-matching model for each pixel.
The RMSD-minimizing search for $f_{\rm eff}$ was performed via globally
shifting the $N_J$ values of each ($n_{\rm H}$, $\chi$) model in steps of
0.01 dex relative to the data, over a range of $\pm$1.0 dex.
(Larger ranges of up to $\pm$10 dex were tested, but did not improve the fits.)
With a minimum RMSD value of 0.050 dex ($\pm$12\%) in
Figure~\ref{fig:pix_vs_mod}, it is obvious that very good
agreements can be found between observed and modeled column densities in
terms of the overall shape of the $N_J$ curve.

As a test case, we fit several regions with fixed values of
$n_{\rm H} = 10^5$ cm$^{-3}$ and $\chi = 5 \times 10^3$.
The median RMSD value on the SR is 0.144 dex,
or a difference of 39\% between data and models, i.e., clearly not fitting
the data within the expected observational uncertainties.
See Table 3 for a listing of results for fixed $n_{\rm H}$ and $\chi$ that
include other regions around the SR.
Owing to the poor fit by a single model of fixed parameters, better
agreement between data and models is expected from expanded ranges of the two
parameters $n_{\rm H}$ and $\chi$.

\subsubsection{Expansion to a Multi-Model Grid}

\begin{figure}
\epsscale{0.6}
\plotone{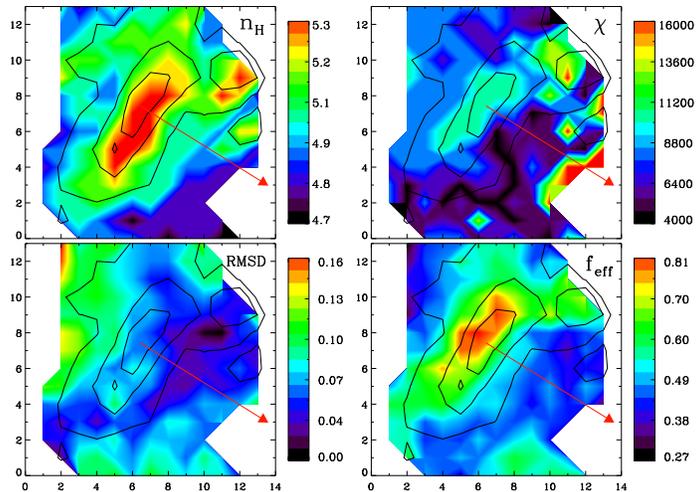}
\caption{Mapping of NGC 2023 with a grid of normal models.
Upper left: logarithmic map of gas density variations.
Upper right: linear map of FUV flux variations.
Lower left: RMSD values are $\leq$0.16 dex over the map.
Lower right: ratio of data to model required to minimize the RMSD.
Each box spans 14 $\times$ 13 pixels, or $\approx 1\arcmin \times 1\arcmin$.
Contours show the 30, 50 and 75\% levels of the total intensity of all five
H$_2$ emission lines (from last panel of Figure~\ref{fig:show_5xem}).
Red arrows extend exactly one half the distance toward HD 37903.}
\label{fig:nor_anyx}
\end{figure}

In order to reveal density and FUV flux variations, and thus take advantage
of the mapping performed by the IRS, we have generated a grid of models
over the $n_{\rm H}$ $vs$ $\chi$ parameter space.
The grid of 49 normal models used here covers 0.6 dex in $n_{\rm H}$ and 0.6
dex in $\chi$ with a step size of 0.1 dex.
It is a subset of a larger grid comprising of 210 models, most of
which did not fit any of the observed $N_J$ curves.
The parameter ranges of $n_{\rm H}$ = 0.5---2 $\times 10^5$ cm$^{-3}$ and
$\chi$ = 4---16 $\times 10^3$
were found to provide much better fits (smaller RMSD) than using a single
model over the entire map.
The results of this multi-model fit are shown in Figure~\ref{fig:nor_anyx}.

In this Figure we see that the highest density values of $2 \times 10^5$
cm$^{-3}$ are found on the SR, as well as on its neighbor, the SSE ridge.
The range of $n_{\rm H}$ over the entire map is in very good agreement with
that given by \cite{Wyrowski2000}.
The $\chi$ map shows an enhancement of the FUV field on the SR of
$\chi = 10^4$, which is $\sim$30\% higher than $7.6 \times 10^3$, the central
value expected according to the distance to NGC 2023 ($\S$ 3.1), but is
nonetheless consistent with the upper 1-$\sigma$ value for the expected $\chi$.
(In fact, our initial comparison with expected $\chi$ values based on
the longer distance scale of 450---500 pc for NGC 2023 had resulted in
model-to-expectation ratio of 2.4$^{+0.3}_{-0.2}$.
Such a very significant discrepancy prompted us to investigate the issue
of the distance to HD 37903 more thoroughly, as reported in $\S$ 2.1).
Taken at face value, a prediction of $\chi$ = 10$^4$ means that HD 37903
could be even closer by 1 sigma than the adopted short distance scale,
i.e., 300, instead of 350, pc away\footnotemark[5].

\footnotetext[5]{This distance of 300 pc has now been confirmed by
\textit{HIPPARCOS}, see end of $\S$ 5 for details.}

Both parameters diminish farther from the SR, except for regions with high
$\chi$ values farther
to the North of the SR (lower right corner) that have smaller (projected)
distance from the exciting star, HD 37903.
The median RMSD value for the entire map area is 0.062 (Table 3).
Note that for the SR this value has improved by a factor of $>$2
relative to the single-model fit.
The highest values of the global shift are also found on the SR, where
the median is $-$0.13 dex, or $f_{\rm eff}$ = 0.74.
Most of the mapped (off-SR) area has a value of $f_{\rm eff} \sim$ 0.49,
hence 1.5 times smaller than the on-SR value.

\subsubsection{Comparing $f_{\rm eff}$ with Previous Results}

The $f_{\rm eff}$-corrected RMSD median of 0.062 shows that
the combination of spatially higher-resolution data from \textit{Spitzer}-IRS
and recent improvements in PDR models appears to yield a very good level of
agreement of 15\% between observations and predictions of H$_2$ emission.
Over the map (Fig.~\ref{fig:nor_anyx}) we find that
$-0.57 \leq$ log $f_{\rm eff} \leq -0.09$, or $0.27 \leq f_+ f_- \leq 0.81$,
a variation by a factor of 3.
Since $f_+ f_- < 1$ for all pixels, $f_-$ is the dominant factor for comparing
models to observations of this region (i.e., $f_+ < 1/f_-$).

In their modeling of the SR, \cite{Draine1996} corrected the data from
\cite{Hasegawa1987} by using a beam filling factor of $f_{\rm B}$ = 1/6,
and employed an inclination correction of $f_\theta$ = 5,
or $\theta \approx 78\degr$.
Their $f_{\rm eff} = f_\theta f_{\rm B}$ = 0.83 is closely
matching the higher values found on our map.
\cite{Draine2000} used the same $f_\theta$ from \cite{Draine1996}, but
$f_{\rm B}$ = 1/1.8 = 0.56, resulting in $f_{\rm eff}$ = 2.8, a value
$>$3 times higher than our map's largest value of 0.81.
\cite{Kaufman2006} used a proxy for the beam factor in the form of
two-phased H$_2$ medium \citep{Steiman1997}, in which the dense gas
from the SR was contributing at a level of 20\%, and thus in effect using
$f_{\rm B}$ = 0.2.
In addition, they used $f_\theta$ = 6 in order to achieve a match between data
and models.
In short, this amounts to using $f_{\rm eff}$ = $f_\theta$$f_{\rm B}$ = 1.2,
a value higher by $\sim$50\% than our map's highest value.
The contributions of multiple PDRs along the line of sight, $f_{\rm P}$,
and the angle of FUV illumination, $f_\phi$, were not explicitly mentioned
by these studies.

We first consider the effects of varying $f_{\rm B}$ while holding the other
factors fixed.
Adopting the assumption of $f_{\rm P} f_\phi$ = 1, and using
$f_\theta$ = 5 on the SR, as per \cite{Draine1996}, we find $f_{\rm B}$ =
$f_{\rm eff}/f_\theta$ = 0.74/5. = 0.148.
However, off the SR, where values of $f_{\rm eff}$ are smaller, i.e.,
$f_{\rm eff}$ = $f_\theta$$f_{\rm B}$ = 0.49 (Table 3), 
the implied beam filling factor would also be smaller,
$f_{\rm B}$ = 0.49/5. = 0.098.
This trend is not likely since away from the SR, the source is more
extended and not as concentrated as the SR, and therefore we expect
$f_{\rm B}$ to become larger (approaching 1), not smaller.
Thus we conclude that it is not $f_{\rm B}$ that is driving down the values
of $f_-$ (and thus of $f_{\rm eff}$) away from the SR.
In the next section we explore the possible 3-D shape of the SR and show how
a simultaneous reduction in both $f_\phi$ and in $f_\theta$ may explain
the smaller off-SR values of $f_{\rm eff}$.

\subsubsection{Clues for 3-D Nebular Structure?}

Our map exhibits $f_{\rm eff}$ below unity, with values decreasing
away from the SR and possibly indicating a dominant reduction in any of
the three factors $f_{\rm P}$, $f_\theta$, or $f_\phi$.
All three factors are related to the 3-D structure of the nebula and
their variability over the map should be helpful in deciphering the
relative configuration between on-SR and off-SR regions.

\begin{figure}
\epsscale{0.5}
\plotone{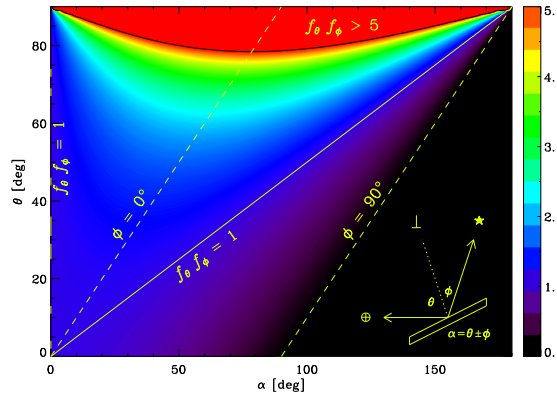}
\caption{Product $f_\theta$$f_\phi$ $\equiv$ $\cos$($\phi$)/cos($\theta$)
is contoured as a function of $\alpha$ and $\theta$.
As depicted in the lower right corner, $\alpha$ is the angle at the source
between HD 37903 ($\star$) and our line of sight ($\earth$),
$\theta$ is the surface inclination of the source, whose normal
is indicated by `$\bot$', and $\phi$ is the angle of FUV incidence.
All values above upper limit of color bar (black contour of
$f_\theta$$f_\phi$ = 5) are mono-colored red.}
\label{fig:theta_and_phi}
\end{figure}

Let us define $\alpha$ as the angle at the source between our line of sight
and the direction of incident FUV radiation from HD 37903.
A contour plot of the product $f_\theta f_\phi$ = cos($\phi$)/cos($\theta$) is
shown in Figure~\ref{fig:theta_and_phi}, based on the relationship
$\alpha = \theta \pm \phi$, or $\phi = \vert\alpha - \theta\vert$.
The lower half of the figure includes the values $f_\theta f_\phi < 1$,
although geometric configurations toward the lower right corner are excluded
owing to the FUV source dipping below the horizon of the PDR face.
There are two lines on the map where $\phi = \theta$, leading to
$f_\theta f_\phi$ = 1.
One is where $\phi = \theta$ = $\alpha$/2., as indicated by the diagonal
line, and the other is where $\alpha = 0\degr$, hence along the ordinate.
Between these two lines and for any given $\theta$, $f_\theta f_\phi$
reaches a maximum when $0\degr \leq \alpha \leq 90\degr$, along the line
$\phi = 0\degr$ where cos($\phi$) = 1 and $\theta = \alpha$.

The impression from Figure~\ref{fig:show_2023} is that HD 37903 is located
at the center of a bubble, and that the SR (as well as other ridges of
intense emission) is marking the irradiated boundaries of the bubble.
We may thus assume that $\alpha$ is not too far from 90$\degr$, and thus the
value of $f_\theta f_\phi$ is expected to be near its maximal value.
Furthermore, $f_\theta$ probably achieves its highest value on the SR,
assuming a near edge-on view of the bubble's boundary.
Away from emission ridges no such narrow features are seen, and
thus moving away from the SR would result in lowering of the value
of $f_\theta f_\phi \rightarrow 1$, as $\phi \rightarrow \theta$. 
For the spherical bubble scenario, we expect that
$45\degr \lesssim \alpha \lesssim 135\degr$, which results in
$0.5 \lesssim f_\phi \leq 1.0$ and thus yielding $f_\theta f_\phi > 1$
for any $f_\theta > 2$.
This constraint is certainly true for $\theta = 78\degr$,
the nominally adopted value from \cite{Draine1996}.

\begin{figure}
\epsscale{0.35}
\plotone{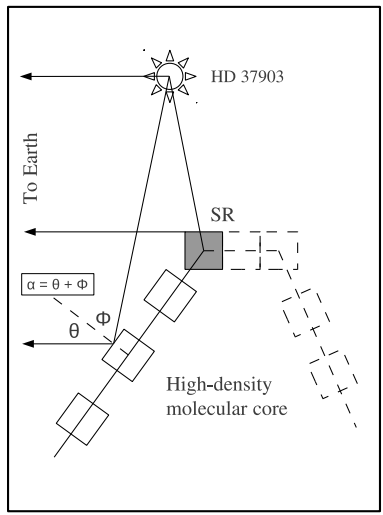}
\caption{Cross-sectional cartoon of the suggested 3-D structure of NGC 2023
in the vicinity of the SR.
The SR sits on top of a quasi-pyramidal high-density molecular core, the
surface of which is made up of parallel cloud ridges that are subject to
reduced levels of FUV irradiation.
This core harbors the heavily obscured formation site of Sellgren D, see
Figures~\ref{fig:show_2023} and \ref{fig:show_irac}.}
\label{fig:cartoon}
\end{figure}

One way to achieve a 3-D configuration such that $f_{\rm eff}$ becomes smaller
with distance from the SR is to assume that the molecular cloud banks have
a quasi-pyramidal cross section.
A cartoon that presents such a configuration is shown in
Figure~\ref{fig:cartoon}.
[An earlier, albeit more simplified, cartoon was given in Figure 4 of
\cite{Field1994}.]
Thus the southern slope of the SR may face our line of sight (with smaller
$\theta$ and $f_\theta \rightarrow 1$), while the surface normal is at
a larger $\phi$ ($f_\phi \rightarrow 0$) from the direction to HD 37903.
In this case, starshine can still hit directly in a normal fashion the top of
the SR that rise toward the star (as can be seen in Figure~\ref{fig:show_2023}),
but the off-SR gas is facing away from HD 37903 and is receiving
slanted, and thus reduced, levels of FUV flux.
In particular, still keeping a range of
$45\degr \lesssim \alpha \lesssim 135\degr$, but assuming a 45$\degr$ turn of
the viewed slab toward our line of sight
(e.g., $\theta = 78\degr - 45\degr = 33\degr$)
allows for a range of $\phi \geq 12\degr$ such that
$0.00 \leq f_\phi \leq 0.98$.
In other words, relative to on-SR values, the off-SR pyramidal slope provides
a combination of smaller $\theta$ and larger $\phi$ that can readily
achieve $f_\theta f_\phi \ll 1$, thus accounting for the observed reduction
in $f_{\rm eff}$ away from the SR. 

In this picture the SR is a manifestation of one ridge of clouds forming
the outer layer of the much larger, denser, and darker molecular cloud.
This cloud was outlined by HCO$^+$ via the millimeter observations of
\cite{Wyrowski2000}, whose Figures 3 and 4 clearly show
that the PDR (H$_2$ and C recombination line emission)
is located on the HD 37903-facing side of the denser structure.
Indeed, the more opaque off-SR sight line may also explain any potential
reduction in $f_{\rm P}$ relative to the on-SR sight line, thus providing a
3-D scenario that simultaneously reduces all three factors
affecting $f_{\rm eff}$.

\subsection{Excitation Diagrams of H$_2$}

\subsubsection{Construction and Interpretation}

While it has been understood since the earliest observations
\citep{Gatley1987,Hasegawa1987} that ro-vibrational emission in NGC 2023 is
dominated by FUV fluorescence, the situation regarding rotational emission from
the $v = 0$ state could depend on both collisional and radiative excitations.
From our models we find that
the S(0), S(1), and S(2) lines are produced at PDR depths where H$_2$-H$_2$
collisions thermalize level populations owing to collisional deexcitation
rates that are much higher than radiative decays.
For the lower $T_{\rm gas}$ involved ($<$300 K), relevant critical densities
of the $J$ = 2, 3, and 4 levels for H$_2$-H$_2$ collisions are 20, 360,
and 5100 cm$^{-3}$, respectively, clearly below the inferred gas density of
$\sim10^5$ cm$^{-3}$.
The S(3) and S(5) lines arise at shallower depths near the PDR surface, where
the gas is warmer and H$_2$-H collisions dominate the excitation.
With respective critical densities for collisions with H (at $\sim$600 K) of
$3 \times 10^4$ and $1.8 \times 10^5$ cm$^{-3}$, the upper levels $J$ = 5
and 7 are fully and marginally thermalized, respectively, by such collisions.
This establishes that the excitation temperatures of the H$_2$ derived
directly from observations are more closely associated with the gas
kinetic temperature achieved through photoelectric heating
than with the FUV pumping rate of the H$_2$.

\begin{figure}
\epsscale{0.5}
\plotone{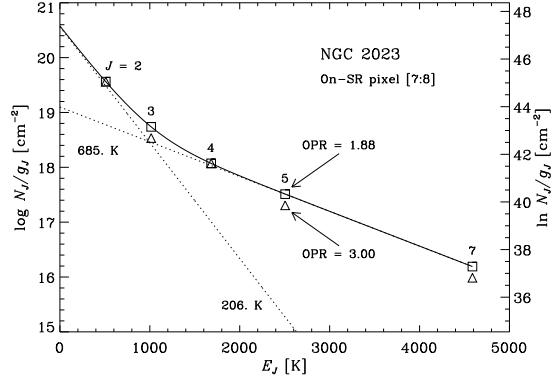}
\caption{Excitation diagram for the on-SR pixel LH[7:8].
Data boxes are $\pm$20\% in vertical extent, owing to IRS uncertainty.
The data were de-zigzagged by shifting the odd-$J$ values for OPR = 3
(triangles) toward the even-$J$ levels, while minimizing the RMSD between
the smooth excitation curve and the data.
This pixel is found to have OPR $\approx$ 1.9, with ``cold'' and ``hot''
$T_{\rm ex}$ of 206 and 685 K, respectively (dashed lines), see also
Figure~\ref{fig:map_t_n0}.}
\label{fig:ex_diag}
\end{figure}

In addition to (model-derived) gas density and FUV intensity, two other facets
of H$_2$ microphysics are the o-H$_2$ to p-H$_2$ ratio \citep{Burton1992}
and the excitation temperatures of rotational-level populations.
Values of the OPR, $T_{\rm ex}$, and ground-state population ($N_0$)
may be directly extracted from excitation diagrams employing
ln $N_J/g_J$ vs $E_J$, as shown in Figure~\ref{fig:ex_diag}.
For gas at LTE and with a single value of $T_{\rm ex}$, the populations obey an
exponential (Boltzmann) distribution so that the excitation function
is a straight line having a slope of $-1/T_{\rm ex}$.
However, when the emitting region includes a range of H$_2$ excitation
temperatures the resulting function will be curved.

Each excitation curve may be approximated by a sum of two asymptotic straight
lines representing two independent single-$T_{\rm ex}$ H$_2$ components.
One is a ``cold'' component with $T_{\rm ex}^{\rm cold} \approx T_{\rm 32}$
and the other is ``hot'' with $T_{\rm ex}^{\rm hot} \approx T_{\rm 75}$.
Extrapolation through $E_1$ and $E_0$ provides estimates of the 
ground state column densities $N_1$ and $N_0$.
However, this method of estimating the H$_2$ column is sensitive
to only those regions that are warm enough to emit in the 
rotational transitions and thus neglects the colder interior gas. In addition, 
owing to the monotonically declining character of the $-1/T_{\rm ex}$ curve,
the straight line fits tend to overestimate $T_{\rm gas}$ (via
$T_{10}$) in the emitting regions.

In the case of H$_2$, the statistical weights $g_J$ include a $2J+1$ factor
from orbital statistics and the nuclear factor of 3 for odd-$J$ levels.
An excitation curve with a misalignment (zigzag) between odd- and even-$J$
populations is the indication that the OPR $\neq$ 3.
The average OPR between observed columns in odd- and even- $J$ states may be
determined by de-zigzagging the curve.
A fitting procedure was accomplished by shifting the odd-$J$ $N$ values toward
the even-$J$ levels, with the resulting de-zigzagged curve of smallest RMSD
providing the observed value of the OPR.

\subsubsection{Mapping the OPR}

Under local thermal equilibrium (LTE) and $T \geq$ 300 K, o-H$_2$ is three
times more abundant than p-H$_2$, hence OPR $\equiv$ 3, owing to nuclear
statistics.
The OPR, which is initially set at the time H$_2$ is
formed on a surface of a dust grain, can be changed over time via collisions
between H$_2$ molecules and other gas constituents including H and H$^+$,
and through accretion and subsequent ejection from grains 
\citep{Burton1992,Sternberg1999,LeBourlot2000}.
Radiative processes in H$_2$ cannot change an original OPR value since
they involve quadrupole transitions of $\Delta J = 0, \pm2$ in the ground
electronic state.
The first evidence for H$_2$ molecules with OPR $\neq$ 3 in the ISM was
provided by \cite{Hasegawa1987}, who observed vibrationally excited H$_2$
with OPR = 1.4---2.0 toward the SR of NGC 2023.
\cite{Sternberg1999} calculated that an OPR value of 1.7 would be observed
in vibrationally excited H$_2$ even from a gas with o-H$_2$/p-H$_2$ = 3, owing
to differences in pumping rates caused by the propensity of o-H$_2$ to self
shield before p-H$_2$.
Thus, definitive evidence of OPR other than 3 in a PDR has remained elusive.
Recent measurements of the OPR in shocked gas, using pure rotational
transitions of H$_2$, show more definitively that the ratio can differ from 3
owing to time dependent effects of the OPR conversion
\citep[e.g.,][]{Neufeld2006,Maret2009,Neufeld2009}.

\begin{figure}
\epsscale{0.6}
\plotone{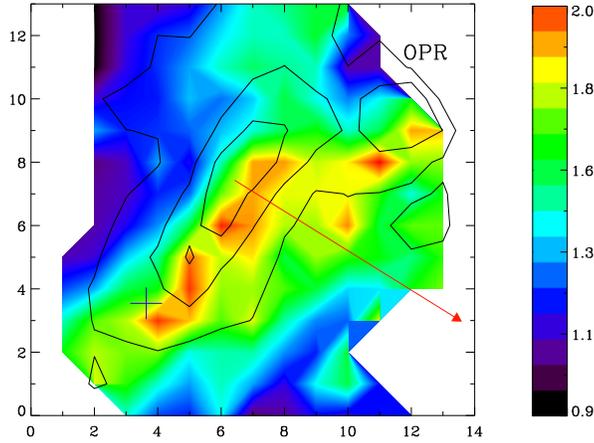}
\caption{Spatial variation of the OPR over the LH field toward NGC 2023.
Box size is 14 $\times$ 13 pixels, or $\approx 1\arcmin \times 1\arcmin$.
Total H$_2$ on-SR intensity in indicated by contour levels of 30, 50 and 75\%,
with the red arrow extending half the distance toward HD 37903
and the location of star C marked by `+'.}
\label{fig:map_opr}
\end{figure}

A de-zigzagging procedure was repeated for all data pixels in the field of
view, yielding the OPR map in Figure~\ref{fig:map_opr}.
First, the global range of OPR variation over the map is 0.9---2.0,
or $\approx$1.5---2.0 inside the 75\% intensity contour on the SR. 
This departure from OPR = 3 is very similar to previous results
\citep{Hasegawa1987,Burton1993,Fleming2010}.
Second, the highest OPR values are not aligned with SR intensity
contours, but are found to be shifted toward the north,
in the direction of the exciting star, HD 37903.
This phenomenon is corroborated by OPR values on the star-facing side
of the neighboring South-Southeastern Ridge.

\begin{figure}
\epsscale{0.6}
\plotone{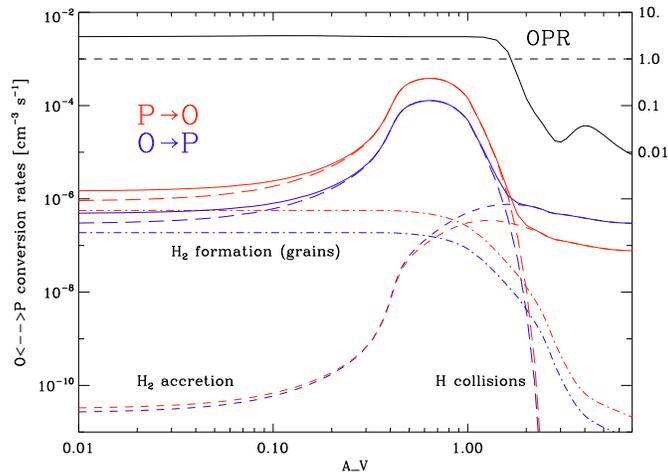}
\caption{Three OPR-controlling processes and their sum are plotted as a
function of depth into the PDR.
Each process is found to dominate o-H$_2$ $\leftrightarrows$ p-H$_2$
conversions over a certain range of $A_V$.
At $A_V \lesssim 1.5$, p-H$_2$ $\rightarrow$ o-H$_2$ domination maintains an
OPR of $\approx$3 (see scale on the right), but deeper into the PDR,
o-H$_2$ $\rightarrow$ p-H$_2$ dominates owing to H$_2$ accretion onto cold dust
and is responsible for a rapid decline of the OPR.
A horizontal dashed line shows the level of OPR = 1.}
\label{fig:opr_rates}
\end{figure}

Modeled OPR values are determined by a variety of conversion processes
between o- and p-H$_2$, with the dominant ones shown in
Figure~\ref{fig:opr_rates}.
Local OPR values are $\gtrsim$3 throughout shallow PDR layers with
$A_V \lesssim 1$, but rapidly decline to $\ll$1 for $A_V \gtrsim 2$.
We performed an identical de-zigzagging analysis of $N_J$ values from the
PDR model output.
The integrated OPR value was found to be 1.8, in agreement with observations.
In other words, the PDR model, with its various routes of
o-H$_2$ $\leftrightarrows$ p-H$_2$, provides a close match to the run of
$T_{\rm ex}$ and H$_2$ level abundances that closely duplicates the observed
OPR in the $J$ = 2---7 states.
This result is important in showing that a steady-state PDR model can naturally
(i.e., without adjustments)
reproduce observed OPR values that are $\neq$ 3, based on integrated columns
through the entire PDR, without recourse to explanations involving time
dependent effects.
We emphasize that owing to collisional domination, the H$_2$ rotational
populations studied here should not be affected by differential self shielding
effects \citep{Sternberg1999}.

\subsubsection{Mapping $T_{\rm ex}$ and $N_0$}

\begin{figure}
\epsscale{0.6}
\plotone{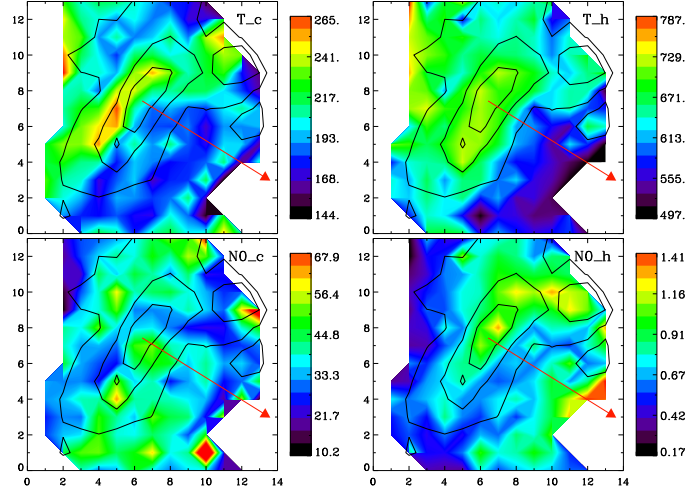}
\caption{Upper panels: mapping of $T_{\rm ex}^{\rm cold}$ (left) and
$T_{\rm ex}^{\rm hot}$ (right) toward the SR, both in K.
Lower panels: mapping of $N_0^{\rm cold}$ (left) and $N_0^{\rm hot}$ (right),
both in 10$^{19}$ cm$^{-2}$.
These 14 $\times$ 13 pixel ($\approx 1\arcmin \times 1\arcmin$) maps are
by-products of the OPR de-zigzagging procedure as demonstrated in
Figure~\ref{fig:ex_diag}.
Red arrows extend half the distance toward HD 37903.}
\label{fig:map_t_n0}
\end{figure}

Values of $T_{\rm ex}$ derived from our \textit{Spitzer} data are
shown in the upper panels of Figure~\ref{fig:map_t_n0}.
Comparing the OPR map of Figure~\ref{fig:map_opr} with the
$T_{\rm ex}^{\rm cold}$ map, we see that the highest $T_{\rm ex}^{\rm cold}$
values are found on the side of the
SR facing away from HD 37903, where OPR values are lower, and vice versa.

Overall, the observed range of OPR values is lower than their LTE values for
the observed range of $T_{\rm ex}^{\rm cold}$.
Specifically, a range of OPR = 0.9---2.0 may be obtained under LTE from
a temperature range of 74---118 K.
All observed $T_{\rm ex}$ values derived from these data, as well as
the $T_{\rm gas}$ model values ($\S$ 4), are $>$144 K and thus
would predict an LTE OPR of $\geq$2.4 over the entire map.
In other words, the H$_2$ has a lower OPR than values corresponding
to all indicators of $T_{\rm gas}$.
(However, as previously remarked, extrapolations of $T_{32}$ to lower-$J$
levels tend to overestimate $T_{10}$, which is not observed via rotational
lines.)
Although this result is derived strictly from observed $N_J$ values, we may
use the model to explain this effect thanks to the very good reproduction of
$N_J$, $T_{\rm ex}$, and the OPR by our PDR model.
We suggest that when $T_{\rm ex}$ and the OPR are calculated from the
integrated columns, $T_{\rm ex}$ is weighted toward $T_{\rm gas}$ in the
line emitting regions, while the OPR is being decoupled from $T_{\rm gas}$
owing to increasing influence of $T_{\rm dust}$.
Indeed, beyond $A_V \approx 1.5$ the OPR is dominated by the
o-H$_2$ $\rightarrow$ p-H$_2$
conversion process via H$_2$ accretion onto cold dust grains
(Figure~\ref{fig:opr_rates}).
These grains have $T_{\rm dust} \ll T_{\rm gas}$, thus
inducing lower integrated OPR values than LTE predictions based on
$T_{\rm gas}$.

The two lower panels of Figure~\ref{fig:map_t_n0} present the derived
$N_0$ populations of ``cold'' and ``hot'' H$_2$, as defined by the
two upper $T_{\rm ex}$ panels.
It is seen that colder H$_2$ is uniformly spread over the map, but that
the hotter H$_2$ population is found concentrated on the bright ridges of
highest intensity.
This picture of correlation between regions of higher $\chi$ values and
hotter H$_2$ is confirmed by the distribution of $T_{\rm ex}^{\rm hot}$
in the same Figure.
Thus observables and derived characteristics that show the distribution of
hotter, excited H$_2$ levels, also track regions with higher FUV
radiation fields.

From a given value of $N_0$ and a single-$T_{\rm ex}$ level population
distribution, the total H$_2$ column density can be estimated.
Using the partition function formula from \cite{Herbst1996} gives
$N_{\rm tot}^{\rm cold} \approx 0.0247 N_0^{\rm cold} T_{\rm ex}^{\rm cold}$ =
$2.4 \times 10^{21}$ cm$^{-2}$.
We may then employ $N_{\rm tot}$ to estimate $n$(H$_2$) 
for an assumed depth along the line of sight.
At the adopted distance of NGC 2023, the SR angular width of $\sim$2$\arcsec$
corresponds to $\sim$$10^{16}$ cm, which closely approximates the model depth
of the warm emitting region.
Employing $f_\theta \approx$ 5 we have
$n$(H$_2$) $\approx N_{\rm tot}^{\rm cold}/5 \times 10^{16} \approx
5 \times 10^4$ cm$^{-3}$.
The corresponding $n_{\rm H} = 2 \times n({\rm H}_2)$ is, of course, higher.
This estimate, which is PDR model independent, provides a strong support
for the case that the SR is a relatively high-density region, and that
our PDR modeling computes very reasonable values for the two parameters
$n_{\rm H}$ and $\chi$.
Note that, just like $N_0^{\rm cold}$, the observational estimates of both
$N_{\rm tot}^{\rm cold}$ and $n$(H$_2$)
are lower limits owing to the underestimation of the unobserved
(but higher-valued) $J$ = 0 and 1 level populations.

\subsection{Atomic Line Emission from Si and Fe}

In $\S$ 2.3 we remarked that the only two PDR-generated atomic emission lines
present in our NGC 2023 spectra belong to [\ion{Si}{2}] and [\ion{Fe}{2}].
The former line at 34.82 $\micron$ is detected from the SR by \textit{Spitzer}
with an average intensity of $2.1 \times 10^{-5}$
erg s$^{-1}$ cm$^{-2}$ sr$^{-1}$.
This is twice the observed intensity estimated by \cite{Kaufman2006}
based on the \textit{ISO} data.

The \cite{Kaufman2006} model prediction for [\ion{Si}{2}] intensity was
$6 \times 10^{-5}$ erg s$^{-1}$ cm$^{-2}$ sr$^{-1}$ based on $n_{\rm H}$ and
$\chi$ from \textit{ISO} H$_2$ observations. 
Their model used a gas phase silicon abundance of Si/H $\sim 1.7 \times 10^{-6}$
for a depletion of $\sim$20 relative to the solar abundance.
The difference between model and observed intensity was attributed to
additional depletion by a factor of $\sim$6 compared to the model value.
There is a strong dependence of model line intensities on both $n_{\rm H}$ and
$\chi$ so that our PDR parameter values, which are based on the \textit{Spitzer}
H$_2$ observations ($\S$ 3.2.2) lead to even higher predicted intensity values.
Figures 1 and 2 of \cite{Kaufman2006} present ($n_{\rm H}$, $\chi$) grids for
normal models, predicting [\ion{Si}{2}] and [\ion{Fe}{2}] emission intensities,
respectively.
We use those grids and depletions for comparison with observed values, and do
not repeat such calculations here since, as far as atomic emission computations
are concerned, essentially the same models used by \cite{Kaufman2006} are
used here.

Our H$_2$ mapping of the SR indicated a log ($n_{\rm H}$, $\chi$) solution
of (5.2, 4.0), for which the predicted [\ion{Si}{2}] intensity
is $\sim$$10^{-3}$ erg s$^{-1}$ cm$^{-2}$ sr$^{-1}$.
Next, this value needs to be corrected for beam filling.
As a rough estimate we use the ratio of the diffraction limit at 35 $\micron$
\citep[8$\farcs$5,][]{Houck2004} over the SR limit of 2$\arcsec$ to get
$\sim$$2 \times 10^{-4}$ erg s$^{-1}$ cm$^{-2}$ sr$^{-1}$.
Consequently, [\ion{Si}{2}] predictions based on our PDR model maps are
$\sim$10 times higher than the observed value.
To bring our observations and model into agreement, the gas-phase Si/H
abundance would have to be $\sim$$1.7 \times 10^{-7}$, i.e., depleted by
a factor of $\sim$200 relative to the solar Si abundance.
Our results are consistent with previous investigations
\citep{Draine2000,Kaufman2006} in finding that a large depletion of gas-phase
Si is required to bring models and observations into agreement.

As for the weaker [\ion{Fe}{2}] line at 25.99 $\micron$, it has 25\%
of the emission intensity of [\ion{Si}{2}], or
$5 \times 10^{-6}$ erg s$^{-1}$ cm$^{-2}$ sr$^{-1}$.
This is the first reported detection of [\ion{Fe}{2}] emission in NGC 2023.
The model-predicted [\ion{Fe}{2}] to [\ion{Si}{2}] intensity ratio is $\sim$0.1,
indicating that the Fe/Si abundance ratio is $\sim$2.5 times higher than
the values adopted in the models.
Since the Si requires further depletion by $\sim$10 we expect the
Fe requires further depletion by $\sim$$10/2.5=4$.
Indeed, the observed [\ion{Fe}{2}] line intensity is well below the predicted
value based on the model abundance of Fe/H = $1.7 \times 10^{-7}$, and requires
a gas phase abundance of Fe/H $\sim 4 \times 10^{-8}$, or $\sim$1/800 times
the solar value.
The depletion is $\sim$4---6 times higher than that observed in diffuse clouds 
\citep{Savage1996} and $\gtrsim$50 times higher than that estimated in several 
PDRs \citep{Okada2008}.
Our anomalously high Fe depletion could be a reflection of the higher value
of on-SR gas density relative to densities probed in previous studies.

\section{Gas Heating Processes}

As was mentioned in the Introduction, previous studies have indicated
the presence of low-$J$ $T_{\rm ex}$ values that seem to be rather high and
thus pose a challenge to models of gas heating in PDR environments.
\cite{Timmermann1996} found $T_{\rm ex}$ $\sim$ 500 K in S140 and were
able to model observed line intensities by employing an initial gas
temperature (at the computational edge of the cloud) of $T_0$ = 1000 K,
as well as cos($\theta$) = 0.1.
\cite{Fuente1999} found a range of $T_{\rm ex}$ $\approx$ 300---700 K in
NGC 7023.
Toward the southern half of NGC 2023, \cite{Fleming2010}
found $T_{\rm ex}$ $\approx$ 500---1400 by fitting H$_2$ excitation curves
with a single (hot) component over an area larger than the one considered here.
Finally, \cite{Habart2011} analyzed \textit{Spitzer} data from a few PDRs with
low-valued FUV fields and found OPR-independent ($\Delta J = 2$) $T_{\rm ex}$
values between 200---750 K toward the northern half of NGC 2023.

Our data toward NGC 2023 provide de-zigzagged values of $T_{\rm ex}$ that
range over 240---700 K for on-SR readings and essentially
overlap all the pure-rotational results for other PDRs mentioned above.
Furthermore, our PDR modeling, which consistently solves for the temperature
structure of the gas for each PDR depth layer, shows a $T_{\rm gas}$ range
of 300---750 K over the H$_2$ line formation region of, e.g.,
$A_V$ $\sim$ 0.5---2.
The close correspondence between $T_{\rm ex}$ and $T_{\rm gas}$ values
calculated by the PDR models confirms that the pure-rotational H$_2$ emission
lines detected and analyzed here are primarily thermally excited by collisions
rather than radiatively excited by FUV photons.

\begin{figure}
\epsscale{0.6}
\plotone{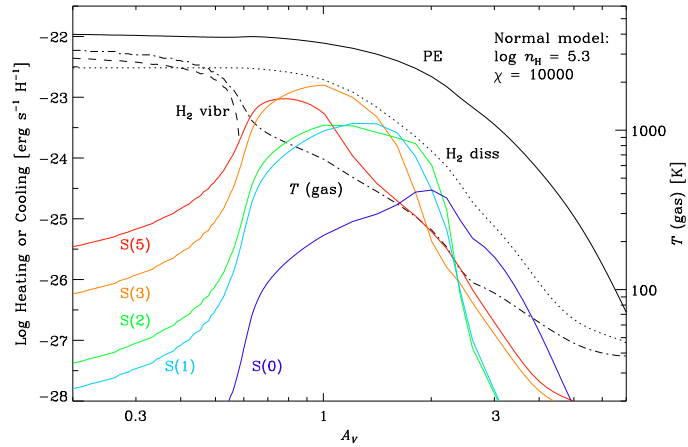}
\caption{Modeled heating processes in the gas for a normal model
with $n_{\rm H} = 2 \times 10^5$ cm$^{-3}$ and $\chi = 10^4$.
Heating is controlled by photoelectric emission from grains
(``PE'', solid curve) throughout the H$_2$ line formation region indicated
by colored emissivity (cooling) curves.
Lesser heating contributions are from H$_2$ dissociation (dotted curve) and
from H$_2$ ro-vibrational de-excitation (dashed curve plotted is net heating
minus cooling).
The dot-dashed curve shows the gas temperature profile, with scale on the
right.}
\label{fig:mod_heating}
\end{figure}

Figure~\ref{fig:mod_heating} shows the depth variations of prominent gas
heating processes for parameter values
$n_{\rm H} = 2 \times 10^5$ cm$^{-3}$ and $\chi = 10^4$.
These processes include grain photoelectric heating as well as heating
via collisional de-excitation of FUV pumped H$_2$, and via H$_2$ dissociation.
Clearly, grain photoelectric heating is dominant throughout the PDR
layers where H$_2$ line emission arises, with other heating processes
contributing $\lesssim$25\% of the heating budget.
The successful reproduction of H$_2$ data toward NGC 2023 implies
that FUV interactions with the dust and gas components of this PDR do not
require an additional energy input in the form of mechanical heating from,
e.g., turbulence or shocks in this source.

\section{Concluding Remarks}

The rich molecular spectrum of H$_2$ provides a rigorous diagnostic tool in
the study of PDRs, the FUV-irradiated envelopes of molecular clouds.
We showed that very good agreement can be obtained between modeled and observed
absolute values of the line intensities of rotationally-excited H$_2$ toward
the SR in NGC 2023.
According to our PDR models, the highest values of H$_2$ emission, which emanate
from the narrow SR, require densities up to $\sim$$2 \times 10^5$ cm$^{-3}$
and radiation fields up to $\sim$$10^4$ times the local Galactic field.
These values are well within the observationally acceptable range and are
consistent with other PDR observables.
The agreement between data and models is a direct result of improved sampling
of the emission owing to the fine spatial resolution of \textit{Spitzer}
instruments, as well as of recent improvements in PDR modeling, including
a more detailed treatment of H$_2$, and an enhanced H$_2$ formation rate
for PDRs.
Our results do not confirm the finding by \cite{Habart2011} of
order-of-magnitude discrepancies between PDR model results and \textit{Spitzer}
H$_2$ data, which could be explained by their observations of
PDRs illuminated by mainly low-FUV fields.
In contrast, for NGC 2023 the FUV field is sufficiently high to dominate 
any non radiative heating that might be present. 

The fact that our model gives a good match to H$_2$ rotational line
intensities, and to their associated run of increasing $T_{\rm ex}$ with $J$,
is a good indication that the model includes an adequate treatment
of heating and cooling processes.
In particular, the dominant heating process via photoelectrons is sufficient
to maintain the correct $T_{\rm gas}$ profile and H$_2$ emission distribution
within the PDR, without additional mechanical sources of heating.
Furthermore, an OPR resulting from H collisions and grain accretion provides 
a value that is $\neq$3 and is matched by fitting both observed and modeled
H$_2$ column densities.
Thus our steady state computation produces the observed ratios among even- and
odd-$J$ states, without the need to artificially adjust the OPR.

According to the PDR model, collisional excitation dominates the
pure-rotational emission lines studied here.
Thus the curvature of rising $T_{\rm ex}$ with $J$ is not the result of 
FUV pumping but of collisional thermalization of H$_2$ levels in tandem with
rising gas temperature within shallower layers of the PDR.
In effect, the small fraction of radiative energy from HD 37903 that 
photoelectrically heats the gas is more important in controlling rotational
level populations via collisions than FUV fluorescence.

Our IRS maps show that the decrease in H$_2$ emission intensity away from
the SR is accompanied by reductions in $n_{\rm H}$, $\chi$, and $f_{\rm eff}$,
the latter being the ratio of data to face-on, beam-filled, 
normally-illuminated model intensity.
The analysis of the four factors that affect the value of $f_{\rm eff}$
indicated that its reduction may be facilitated by a combined reduction
in inclination and incidence factors, which are countered by a more
modest increase in the value of the beam-filling factor.
Despite the difficulty of constraining all factors, the observations were shown
to be consistent with a previously \citep{Field1994} suggested 3-D structure
of the region around the SR, namely a quasi-pyramidal molecular cloud
towering above the FUV-carved cavity toward HD 37903.

The successful interplay between PDR observations and theory supports
the description of NGC 2023 as a \textit{par excellence} example of
a photodissociation region.
This is especially significant because the results presented here require the
shorter distance scale for HD 37903.
Indeed, a very recent re-examination of the SIMBAD database on 2011 May 20 at
12:20 EDT revealed a newly published revision of the \textit{HIPPARCOS}
parallaxes based on \cite{Leeuwen2007}.
One of these parallaxes provides a new distance to HD 37903 of
300$^{+110}_{-60}$ pc, predicting an on-SR $\chi$ of
$(1.03^{+0.58}_{-0.48}) \times 10^4$, which is in extremely good agreement with
our PDR modeling results.

\acknowledgements

YS thanks Dr. Fr\'{e}d\'{e}ric Galliano, whose table, chair, and a very useful
suite of start-up IDL routines were inherited by YS upon arrival at UMCP.
This work was supported in part by NASA through an award issued by
JPL/Caltech and is based on observations made with the \textit{Spitzer} Space
Telescope, which is operated by the Jet Propulsion Laboratory, California
Institute of Technology under NASA contract 1407.
YS and MGW were supported in part by a NASA Long Term Space Astrophysics
Grant NNG05G64G.
MJK and MC were supported by NSF grant PHY-0552964 for the REU program in
Physics and Astronomy at San Jose State University.
This research has made use of the SIMBAD database, operated at CDS, Strasbourg,
France.

Facilities: \facility{Spitzer (IRS, IRAC)}, \facility{ESO: VISTA}, \facility{HST (ACS)}

\clearpage

\begin{deluxetable}{ccccccccc}
\tablecolumns{9}
\tablewidth{0pt}
\tabletypesize{\footnotesize}
\tablecaption{Log of \textit{Spitzer} Observations of NGC 2023}
\tablehead{\colhead{Module} &\colhead{Pixel} &\colhead{Slit} &\colhead{AOR} &\colhead{Object} &\colhead{Date} &\colhead{$\alpha$(J2000)} &\colhead{$\delta$(J2000)} &\colhead{Exposures} \\
  &\colhead{($\arcsec$)} &\colhead{($\arcsec\times\arcsec$)} & & & & & &(\# $\times$ s) }
\startdata
SH  &2.26  &4.7 $\times$ 11.3  &14033920  &NGC 2023  &2006/03/19  &05:41:37.63  &$-$02:16:42.6  &144 $\times$ 30  \\
LH  &4.46  &11.1 $\times$ 22.3 &14034176  &NGC 2023  &2006/03/19  &05:41:37.63  &$-$02:16:42.6  &30 $\times$ 60  \\
SL  &1.85  &3.7 $\times$ 57    &17977856  &NGC 2023  &2007/10/08  &05:41:37.63  &$-$02:16:42.6  &54 $\times$ 28  \\
SL  &1.85  &3.7 $\times$ 57    &17978112  &Sky bkgd  &2007/10/08  &05:40:26.21  &$-$02:54:40.3  &4 $\times$ 28  \\
\enddata
\end{deluxetable}

%\clearpage

\begin{deluxetable}{cccccccc}
\tablecolumns{8}
\tablewidth{0pt}
\tabletypesize{\footnotesize}
\tablecaption{GAUSSFIT Parameters\tablenotemark{a} and their Uncertainties\tablenotemark{b}}
\tablehead{\colhead{Line} &\colhead{$\lambda_{\rm rest}$} &\colhead{$A_0$\tablenotemark{c}} &\colhead{$A_1$} &\colhead{$A_1-\lambda_{\rm rest}$} &\colhead{$A_2$} &\colhead{$R$\tablenotemark{d}} &\colhead{$\Sigma A_0$\tablenotemark{c}} \\
 &\colhead{($\micron$)} &\colhead{(10$^{-5}$)} &\colhead{($\micron$)} &\colhead{($\micron$)} &\colhead{($\micron$)}  & &\colhead{(10$^{-5}$)}}
\startdata

 & &\multicolumn{6}{c}{Map Median (off-SR)} \\
\cline{3-8}
S(0) &28.2188 &1.35(8) &28.225(1)  &$+$0.006 &0.015(1)  &804(65) &2.3(2)  \\
S(1) &17.0348 &5.84(9) &17.0378(2) &$+$0.003 &0.0106(2) &680(12) &11.6(3) \\
S(2) &12.2786 &6.9(1)  &12.2801(1) &$+$0.001 &0.0071(1) &736(14) &13.5(4) \\
S(3) &9.6649  &9.2(2)  &9.663(1)   &$-$0.002 &0.053(2)  &78(2)   &19.6(8) \\
S(5) &6.9095  &4.1(5)  &6.905(3)   &$-$0.005 &0.030(4)  &98(12)  &9(2)    \\
\tableline
R.u.\tablenotemark{e} & &2---12\% &0.001---0.04\% &      &2---12\%   &2---12\% &3---18\% \\
\\
 & &\multicolumn{6}{c}{Single Pixel [7:8] (on-SR)} \\
\cline{3-8}
S(0) &28.2188 &1.7(1)  &28.2231(7) &$+$0.004  &0.014(1)  &844(64) &2.8(3)  \\ 
S(1) &17.0348 &14.8(2) &17.0380(2) &$+$0.003  &0.0105(2) &688(13) &29.4(7) \\
S(2) &12.2786 &18.8(2) &12.2798(1) &$+$0.0008 &0.0068(1) &772(13) &33.8(7) \\
S(3) &9.6649  &40.8(4) &9.6689(5)  &$+$0.004  &0.0511(6) &80.3(9) &84(1)   \\
S(5) &6.9095  &22(1)   &6.912(2)   &$+$0.002  &0.031(2)  &95(6)   &54(4)   \\
\tableline
R.u.\tablenotemark{e} & &1---6\%  &0.001---0.02\% &       &1---8\%    &1---8\%  &1---9\%  \\
\enddata
\tablenotetext{a}{Each line is fitted with $I(\lambda) = A_0exp(-0.5[\lambda-A_1]^2/A_2^2)+A_3+A_4\lambda$;
the continuum (last two terms) is subtracted from the fit.}
\tablenotetext{b}{Uncertainties for the last digits are in parentheses.}
\tablenotetext{c}{Units for $A_0$ are erg s$^{-1}$ cm$^{-2}$ sr$^{-1}$; $\Sigma A_0$ = integrated line intensity,
with conservative uncertainty taken from $A_0$ \& $A_2$ in quadratures.}
\tablenotetext{d}{$R \equiv \lambda_{\rm rest}/(2.355A_2)$ is the spectral resolution.}
\tablenotetext{e}{R.u. $\equiv$ Relative uncertainty.}
\end{deluxetable}

%\clearpage

\begin{deluxetable}{lcccrc}
\tablecolumns{6}
\tablewidth{0pt}
\tabletypesize{\small}
\tablecaption{Normal Model Mapping of NGC 2023-South}
\tablehead{\colhead{Map region} &\colhead{$n_{\rm H}$\tablenotemark{a}} &\colhead{$\chi$} &\colhead{RMSD} &\colhead{$f_{\rm eff}$\tablenotemark{b}} &\colhead{$\chi$/$n_{\rm H}$} \\
  \colhead{(\# of pixels)} &\colhead{(dex)} & &\colhead{(dex)} & &}
\startdata
  &\multicolumn{5}{c}{Model: log $n_{\rm H}$ = 5.0; log $\chi$ = 3.7} \\
  \cline{2-6}

Global (170)    &5.0  &5000.  &0.104  &0.50  &0.05 \\
On-SR (15)      &5.0  &5000.  &0.144  &1.07     &0.05 \\
Off-SR-S (9)    &5.0  &5000.  &0.104  &0.60  &0.05 \\
Off-SR-N (12)   &5.0  &5000.  &0.134  &0.37  &0.05 \\
\\
  &\multicolumn{5}{c}{Grid: log $n_{\rm H}$ = 4.7---5.3; log $\chi$ = 3.6---4.2} \\
  \cline{2-6}

Global (170)    &5.0  &8000.  &0.062  &0.49  &0.08  \\
On-SR (15)      &5.2  &10000. &0.062  &0.74  &0.06  \\
Off-SR-S (9)    &5.1  &6000.  &0.075  &0.48  &0.05  \\
Off-SR-N (12)   &4.9  &6000.  &0.064  &0.49  &0.08  \\

\enddata
\tablecomments{Values are sample medians.}
\tablenotetext{a}{Number density units are cm$^{-3}$.}
\tablenotetext{b}{Ratio of data/model.}
\end{deluxetable}

\clearpage

%\clearpage

%\clearpage

%\clearpage

%\clearpage

%\clearpage

%\clearpage

%\clearpage

%\clearpage

%\clearpage

%\clearpage

%\clearpage

%\clearpage

\end{document}